\documentclass[conference,twocolumn,10pt,a4paper]{IEEEtran}

\ifCLASSINFOpdf

\else

\fi

\usepackage{graphicx} 
\usepackage{booktabs} 
\usepackage{amsmath}
\usepackage{amsfonts}
\usepackage{amssymb}
\usepackage{epstopdf}
\newcounter{MYtempeqncnt}
\usepackage{dblfloatfix}
\usepackage[caption=false,font=footnotesize]{subfig}
\usepackage{color}

\usepackage{cite}

\begin{document}
%
\title{Power Talk: How to Modulate Data over a DC Micro Grid Bus using Power Electronics}

\author{\IEEEauthorblockN{Marko Angjelichinoski, \v Cedomir Stefanovi\' c, Petar Popovski}
\IEEEauthorblockA{Department of Electronic Systems\\
Aalborg University\\
Email: \{maa,cs,petarp\}@es.aau.dk}
\and
\IEEEauthorblockN{Hongpeng Liu, Poh Chiang Loh, Frede Blaabjerg}
\IEEEauthorblockA{Department of Energy Technology\\
Aalborg University\\
Email: \{hpl,pcl,fbl\}@et.aau.dk}}

\maketitle

\begin{abstract}

We introduce a novel communication strategy for DC Micro Grids (MGs), termed \emph{power talk}, in which the devices communicate by modulating the power levels in the DC bus.
The information is \emph{transmitted} by varying the parameters that the MG units use to control the level of the common bus voltage, while it is received by processing the bus measurements that units perform.
This communication is challenged by the fact that the voltage level is subject to random disturbances, as the state of the MG changes with random load variations.
We develop a corresponding communication model and address the random voltage fluctuations by using coding strategies that transform the MG into some well-known communication channels.
The performance analysis shows that it is possible to mitigate the random voltage level variations and communicate reliably over the MG bus.

\end{abstract}


%
\IEEEpeerreviewmaketitle

\section{Introduction}
\label{sec:intro}

The future smart grid is envisioned as a network of micro grids (MGs), localized clusters of distributed energy resources (DERs) and loads that are connected in parallel to a common MG bus(es) through power electronic interfaces \cite{ref0,ref1}.
Advanced control systems, supported with communication technologies, are essential for proper operation of MGs \cite{ref2,ref2_new}.
Unlike traditional power grids, where the feedback of the control system is enabled with external communication network, recent advances in MG control architectures suggest that the operation of the MG should not be critically linked to external communication channels \cite{ref3}.
It is, therefore, practical to consider using the MG itself, i.e., electronic converters, power lines and other equipment, to carry out the communication.
In this sense, we use the term \emph{power talk} to denote potential techniques that modulate the power-related parameters of a power grid in order to send communication signals.

\begin{figure}[t]
\centering
\includegraphics[scale=0.3]{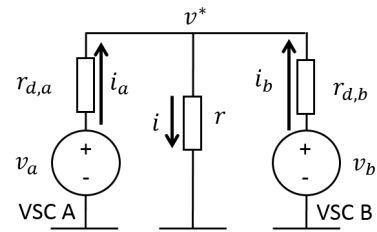}
\caption{Simple DC MG with two VSC units.}
\label{Fig3}
\end{figure}

\begin{figure}[t]
\centering
\includegraphics[scale=0.5]{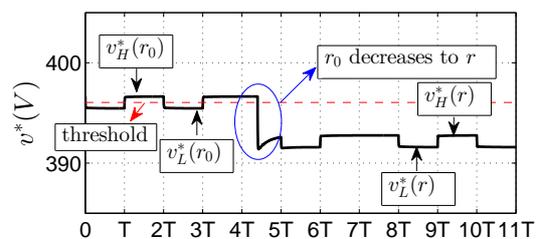}
\caption{The effects of load change.}
\label{Fig6}
\end{figure}

The pivotal idea of power talk can be illustrated through a basic example of DC MG in steady-state, shown in Fig.~\ref{Fig3}.
The MG consists of two units whose power electronic converters control the voltage level of the bus using droop technique, described in Section~\ref{sec:droop}.
The units are denoted with VSC~A and VSC~B, their droop parameters are denoted with $v_a$, $r_{d,a}$ and $v_b$, $r_{d,b}$, respectively, and they are connected in parallel to a pure resistive load $r$.
As the load changes, the power electronics of the units reset the bus voltage $v^*$ to accommodate the new power demand.
Now, assume that VSC~A wants to communicate to VSC~B some information, and, to do so, \emph{changes} the reference voltage $v_a$.
Being a function of $v_a$, the bus voltage $v^*$ will also change and this change will be observed by VSC~B.
In the simplest case of binary signaling, we can use two reference voltages: $v_{a,H}$ to transmit `1' and $v_{a,L} < v_{a,H}$ to transmit `0'.
However, the bus voltage is also a function of the load $r$, which changes randomly and alters the level of $v^*$  in an unpredictable manner, thus representing a form of random noise and thereby main communication challenge in the proposed system.
Fig.~\ref{Fig6} depicts recording of a bit-stream transmission over MG bus using the proposed approach\footnote{
Fig.~\ref{Fig6} and Fig.~\ref{Fig9} have been obtained by simulating the system shown on Fig.~\ref{Fig3} using PLECS\textregistered{} (Piecewise Linear Electrical Circuit Simulation) integrated with Simulink\textregistered{}.}, showing how a load change, which takes place in interval $[4T, 5T]$, affects the bus voltage levels and impairs the detection at the receiver.

The objective of this paper is to introduce power talk mechanism for the model depicted in Fig.~\ref{Fig3} and to investigate methods for achieving a reliable communication in presence of unpredictable load variations in the system.
Specifically, we first investigate the potential for varying bus parameters, such that the operational MG constraints are not violated.
We proceed by introducing the notion of signaling space that can be used to design symbol constellations.
We show that the unpredictable load changes can be addressed by using communication strategies such as pilot transmission and line coding, which effectively creates communication channels that are equivalent to some well-known channels.
Finally, we note that the performance of the proposed communication schemes was tested using a real-time PLECS$^{\textregistered{}}$ MG simulator integrated with Simulink$^{\textregistered{}}$.

The communication setup considered in the paper is a non-standard one, and to the best of our knowledge, has not been considered in the existing literature.
Particularly, the nature of the dominant ``noise'' in the system, created by unpredictable load changes, leads to communication-system design issues that are atypical in a more general sense.
Further, we note that related concept of communicating through variations of the common bus parameters (voltage level in DC and/or frequency in AC systems) termed \emph{bus signaling}, has appeared in \cite{ref4}.
However, the bus signaling schemes are confined to the context of enabling decentralized control using fixed voltage and/or frequency thresholds \cite{ref4,ref5,ref6}, or inserting coordination signals on different frequencies in the bus \cite{ref10}, in which the system takes predefined actions when the values of the bus parameters cross over a threshold or a coordination signal on a predefined frequency is received.
On the other hand, power talk assumes a general concept of information transfer over the MG bus; in this sense, bus signaling can be seen as its specific instance.

The rest of the paper is organized as follows.
Section~\ref{sec:droop} briefly discusses droop control in DC MG systems.
Section~\ref{sec:bus_signaling} presents the power-talk channel model and investigates the signaling space.
Section~\ref{sec:binary} provides insights on designing communication strategies and signaling waveforms to mitigate the effect of load changes.
Section~\ref{sec:analysis} and Section~\ref{sec:evaluation} present the related error probability analysis and performance evaluation, respectively.
Section~\ref{sec:conclusions} concludes the paper.

\section{Droop Control in Micro Grids}
\label{sec:droop}

MGs can operate in grid-connected, or in islanded mode (i.e., disconnected from the rest of the grid). 
In islanded mode, the MG units themselves control and maintain the voltage level of the common bus, through their electronic converters operating as Voltage Source Converters (VSC).
We assume that the VSCs employ droop control mechanism, a standard method for decentralized power sharing in islanded MGs \cite{ref3}.
In DC systems, the droop control consists of inserting a Virtual Resistance (VR), also known as droop slope, into the feedback of the power sharing loop.
Then, the output voltage of $k$-th VSC unit, denoted as $v_{o,k}$, is given by the following steady-state law:
\begin{equation}\label{eq.1}
v_{o,k}^*=v_k-r_{d,k}i_k,
\end{equation}
where $v_k$ and $r_{d,k}$ are the droop parameters, i.e., the nominal reference voltage and the VR, and $i_k$ is the output current of the VSC unit.
As the output current varies with the variable power consumption by the loads, the VSC units adjust its output voltage to meet the new power demand.
The voltage references of the VSC units are usually set to the value of the nominal DC bus voltage.
The VRs are designed to enable adequate power sharing among the units in the system, which is commonly performed in a proportional manner based on the power rating of the specific unit.
Assuming that the current rating of $k$-th VSC unit is $I_{k,max}$ and the maximum output voltage drop $v_k-V_{min}$, the droop slope that enables proportional power sharing is $r_{d,k}=\frac{v_k-V_{min}}{I_{k,max}}$.

\section{Power Talk in DC Micro Grids}
\label{sec:bus_signaling}



\subsection{Model of the Communication System}

\begin{figure}[t]
\centering
\includegraphics[scale=0.2]{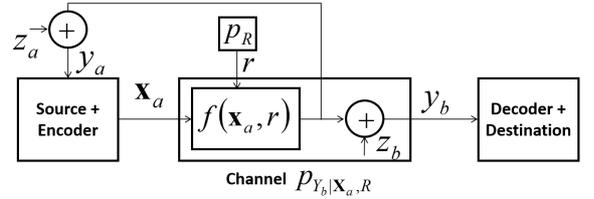}
\caption{Model of the MG as a communication system.}
\label{Fig4}
\end{figure}

To capture the essence of the proposed strategy, we focus on one-way communication from VSC~A to VSC~B, see Fig.~\ref{Fig3}.
In general, both $v_a$ and $r_{d,a}$ of VSC~A can be varied in order to transmit the information to VSC~B.
Thus, the vector:
\begin{equation}\label{eq.2}
\mathbf{x}_a = (v_a,r_{d,a}) \in\mathbb{R}^2,
\end{equation}
is the input to the communication channel.
The bus voltage $v^*$ changes as a function of the channel input:
 \begin{align}
 \label{eq:f}
 v^* = f(\mathbf{x}_a,r).
 \end{align}
The voltage that VSC~B observes (i.e., measures) is:
\begin{equation}\label{eq.3}
y_b = v^* + z_b = f(\mathbf{x}_a,r) + z_b,
\end{equation}
where $z_b\sim\mathcal{N}(0,\sigma^2)$ is the observation noise.
VSC~B through $y_b$  decides which symbols were transmitted by VSC~A.
The physical MG system also provides an immediate and zero-cost feedback available at the transmitter, i.e., VSC~A also observes the bus voltage:
\begin{align}
\label{eq:t}
y_a = v^* + z_a = f(\mathbf{x}_a,r) + z_a,
\end{align}
where $z_a\sim\mathcal{N}(0,\sigma^2)$ is the noise at the transmitter. 
The above communication model is summarized in Fig. \ref{Fig4}.
As expressed in \eqref{eq.3}, the main communication impairments are: 1) random load changes of $r$ and 2) noise of the measurement/observation $z_b$.
The variance of the noise is typically $\sigma^2 << 1 \, \text{V}^2$, thus it can be assumed that the impact of the noise can be neglected.
In the rest of the paper we focus only on the impact of random load changes.

\begin{figure*}[!b]
\normalsize
\hrulefill
\vspace{4pt}
\setcounter{MYtempeqncnt}{\value{equation}}
\setcounter{equation}{9}
\begin{align}\label{eq.11}
\mathcal{D}_{v} & =\left\{(v_a,r_{d,a}) \in\mathbb{R}^2:r_{d,a}\bigg(\frac{V_{min}}{R_{min}}+\frac{V_{min}-v_b}{r_{d,b}}\bigg)+V_{min}\leq v_a\leq r_{d,a}\bigg(\frac{V_{max}}{R_{max}}+\frac{V_{max}-v_b}{r_{d,b}}\bigg)+V_{max}\right\}, \\
\label{eq.12}
\mathcal{D}_{i_a} & =\left\{(v_a,r_{d,a}) \in\mathbb{R}^2:\frac{\frac{v_b}{r_{d,b}}}{\frac{1}{r_{d,b}}+\frac{1}{R_{max}}}\leq v_a\leq r_{d,a}I_{a,max}+\frac{I_{a,max}+\frac{v_b}{r_{d,b}}}{\frac{1}{R_{min}}+\frac{1}{r_{d,b}}}\right\} .
\end{align}
\setcounter{equation}{\value{MYtempeqncnt}}
\end{figure*}

We model the load $r$ as state of the channel, as suggested in Fig. \ref{Fig4}, assuming that it changes slowly with respect to the rate of communication.
In order to deal with random load variations, a feasible communication strategy would be to use pilot symbols to estimate $r$.
The estimation would require the receiver to have full knowledge of the system configuration, expressed through a function $f$ in \eqref{eq:f}.
However, in practice it is  usually difficult to determine or estimate $f$ accurately (using Thevenin equivalent) as the system becomes larger and more complex.
These observations motivate the development of a more robust communication strategy that does not require precise knowledge of the load, as elaborated in Section~\ref{sec:binary}.
Finally, to model the function $f$, we assume that the time period between two signaling instances is long enough so that the system reaches a steady-state.
Using Kirchhoff's laws, the DC bus level for the basic MG example is given by:
\begin{equation}\label{therip}
v^*=f(\mathbf{x}_a,r)=\frac{\alpha(\mathbf{x}_a)}{\beta(\mathbf{x}_a)+r^{-1}},
\end{equation}
where $\alpha(\mathbf{x}_a)=v_ar_{d,a}^{-1}+v_br_{d,b}^{-1}$ and $\beta(\mathbf{x}_a)=r_{d,a}^{-1}+r_{d,b}^{-1}$. 

\subsection{Communication under Constraints: The Signaling Space}

\begin{figure}[t]
\centering
\includegraphics[scale=0.5]{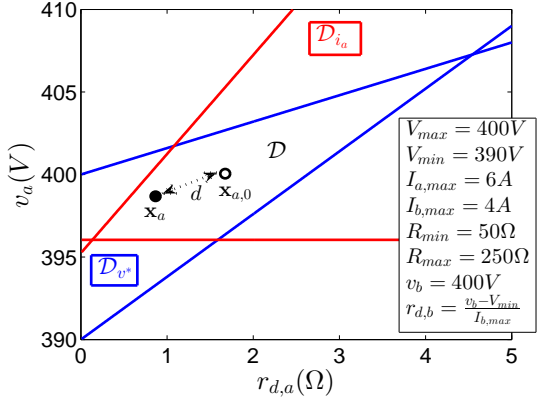}
\caption{The signaling space $\mathcal{D}$ for the system in Fig.~\ref{Fig3} and constraint set $\mathcal{C}$, see (\ref{eq.10}).}
\label{Fig5}
\end{figure}

The MG as a power supply system imposes operational constraints in order to meet predefined power quality demands; we denote the set of constraints with $\mathcal{C}$.
From communication perspective, $\mathcal{C}$ defines the signaling space as the set of all input symbols $\mathbf{x}_a$ that can be used for signaling, such that the predefined voltage and/or current constraints are not violated, irrespective of the changes of the power demand.
Denote the signaling space with $\mathcal{D}$, where:
\begin{align}\label{eq.9}
\mathcal{D} = \left\{\mathbf{x}_a = (v_a,r_{d,a}) \in\mathbb{R}^2:\;\mathcal{C}\right\}.
\end{align}
We proceed by illustrating the signaling space for the system under inspection, as shown in Fig. \ref{Fig3}.
Assuming that the load varies within a predefined interval $r\in[R_{min},\;R_{max}]$ leads to the following set of bus voltage and output current constraints:
\begin{align}\label{eq.10}
\mathcal{C}=
\left\{V_{min}\leq v^*\leq V_{max},0\leq i_a\leq I_{a,max}\right\},
\end{align}
The corresponding signaling space is $\mathcal{D}=\mathcal{D}_{v^*}\cap\mathcal{D}_{i_a}$, where $\mathcal{D}_{v^*}$ and $\mathcal{D}_{i_a}$ are generated by each constraint separately, accounting that $r\in[R_{min},R_{max}]$.
$\mathcal{D}_{v^*}$ and $\mathcal{D}_{i_a}$ are given in (\ref{eq.11}) and (\ref{eq.12}), respectively.
Fig.~\ref{Fig5} depicts $\mathcal{D}$ under a representative example set of values of the MG parameters.
The bounding lines of the signaling space represent the output voltage and current response when $r=R_{min}$ or $r=R_{max}$.
They can be easily obtained by estimating the parameters of the equivalent Thevenin circuit as seen by VSC A at maximum and minimum load, which can be done offline in the training phase, without explicitly knowing the configuration of the system.

\begin{figure}[t]
\centering
\includegraphics[scale=0.5]{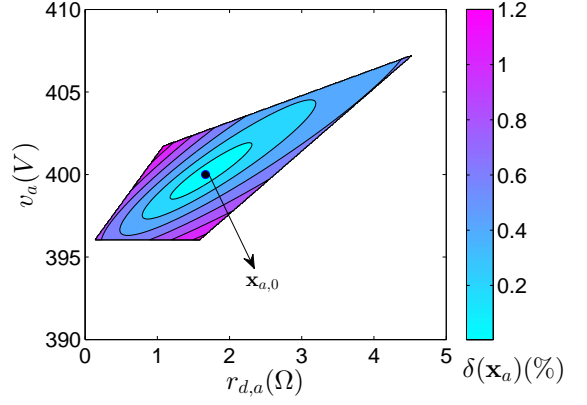}
\caption{The average relative power deviation $\delta$ for symbols from the signaling space ${D}$.}
\label{Fig_diss}
\end{figure}

For a given load $r$, each symbol $\mathbf{x}_{a}\in\mathcal{D}$ produces a different output power $P(\mathbf{x}_{a},r)=\frac{(v^*)^2}{r}$.
To account for this effect, we introduce the constellation design parameter $\delta(\mathbf{x}_{a})$ as the average relative power deviation w.r.t. the nominal operating point $\mathbf{x}_{a,0}$:
\begin{equation}\label{latest1}
\delta(\mathbf{x}_{a})=\frac{\sqrt{\mathbb{E}_{R}\left\{[P(\mathbf{x}_{a},r)-P(\mathbf{x}_{a,0},r)]^2\right\}}}{\mathbb{E}_{R}\left\{P(\mathbf{x}_{a,0},r)\right\}}
\end{equation}
where the averaging is performed over $r$.
The goal is to design symbol constellations with low $\delta$.
Intuitively, the value of $\delta$ is determined by the distance $d=||\mathbf{x}_{a}-\mathbf{x}_{a,0}||_{2}$, see Fig.~\ref{Fig5}.
Fig.~\ref{Fig_diss} shows $\delta$ (in \%) for symbols from $\mathcal{D}$, assuming the uniform distribution $R\sim\mathcal{U}[R_{min},R_{max}]$. 
As expected, $\delta$ grows with the distance $d$; 
the increase in $\delta$ is non-isotropic, as a result of the steady-state model (\ref{therip}), which is non-linear in $\mathbf{x}_a$ as well as $r$.
In practical systems with strict constraints that strive to minimize the power deviation introduced with power talk, one should choose symbols in the close vicinity of the pilot, making the region of small $\delta$ of practical importance.


\section{Binary Power Talk }
\label{sec:binary}

\begin{figure}[t]
\centering
\includegraphics[scale=0.5]{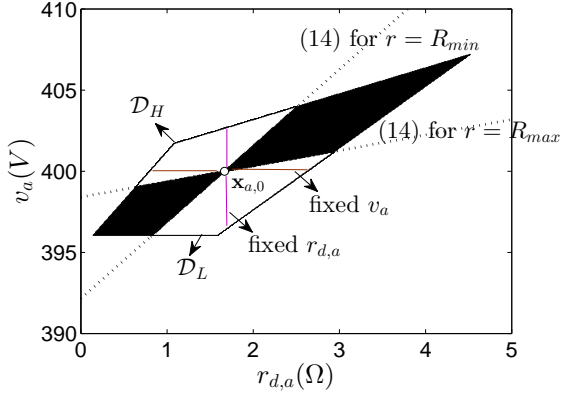}
\caption{Designing binary symbol constellation: $v_{H}^*>v_0^*>v_{L}^*$.}
\label{Fig7}
\end{figure}

In this section we investigate some simple binary power-talk protocol designs.
We assume that the time is slotted in intervals of duration $T$ and that VSCs are slot-synchronized. 
A slot represents a single bit and is henceforth referred to as a bit interval.
The actual communication is performed using constellation symbols $\mathbf{x}_a$ and the interval in which a symbol is inserted is referred to as symbol interval, whose duration is denoted by $T_S$.
A single bit can be communicated via one or multiple symbols, thus the bit interval can consist of one or multiple symbol intervals.
Further, $T_S$ is chosen such that the bus reaches a steady state within a symbol interval, which depends on the control bandwidth of the MG.
Finally, we assume that the load changes occur sparsely in time, such that there could be a maximum of a single load change within a bit interval; this assumption has also been verified in the MG simulator used to assess the proposed communication schemes.

\subsection{Transmitter Design}

Let the \emph{pilot symbol} $\mathbf{x}_{a,0}$ contain the control parameters that VSC~A applies in a nominal operation.
The pilot produces the referent bus voltage $v_0^*=f(\mathbf{x}_{a,0},r)$, establishing the threshold level used by the receiver to decide which of the binary information symbols has been transmitted and thus removing the uncertainty related to the unknown value of $r$.
The pilot is sent at the beginning of the communication and every time the transmitter observes a change in $r$ that leads to incorrect decisions at the receiver; recall that there is an immediate feedback available at the transmitter, see \eqref{eq:t}.

The information symbols are denoted as $\mathbf{x}_{a,H}\in\mathcal{D}$ and $\mathbf{x}_{a,L}\in\mathcal{D}$, corresponding to bus voltage levels $v_{H}^*=f(\mathbf{x}_{a,H},r)$ and $v_{L}^*=f(\mathbf{x}_{a,L},r)$.
Although the function $f$ is not known in detail, the symbols can be chosen within the signaling space to satisfy the following conditions:
\addtocounter{equation}{2}
\begin{align}
\label{eq.5}
v_{H}^* & =f(\mathbf{x}_{a,H},r) > v_0^*,\\
\label{eq.6}
v_{L}^* & =f(\mathbf{x}_{a,L},r) < v_0^*.
\end{align}
Under the model (\ref{therip}), the above conditions translate to:
\begin{align}\label{eq.13}
v_{a} \underset{L}{\overset{H}{\gtrless}}r_{d,a}\frac{v_{a,0}\big(\frac{1}{r_{d,b}}+\frac{1}{r}\big)-\frac{v_b}{r_{d,b}}}{1+\frac{r_{d,a,0}}{r_{d,b}}+\frac{r_{d,a,0}}{r}}+\frac{\frac{v_{a,0}}{r_{d,a,0}}+\frac{v_b}{r_{d,b}}}{\frac{1}{r_{d,a,0}}+\frac{1}{r_{d,b}}+\frac{1}{r}}.
\end{align}
To meet requirements \eqref{eq.5} and \eqref{eq.6} irrespective of the load variations, the symbols have to satisfy the conditions in \eqref{eq.13} for any $r\in[R_{min},R_{max}]$.
This leads to the two disjoint regions $\mathcal{D}_{H}$ and $\mathcal{D}_{L}$ that compose $\mathcal{D}$, depicted in Fig.~\ref{Fig7}. 
In other words, if $\mathbf{x}_{a,H}\in\mathcal{D}_{H}$ and $\mathbf{x}_{a,L}\in\mathcal{D}_{L}$, then $v_L^*<v_0^*<v_H^*$, irrespective of the value of $r$.
Fig.~\ref{Fig7} also depicts two simple symbol constellation designs: 1) i.e. place the symbols on horizontal line through $\mathbf{x}_{a,0}$ such that $v_{a,H}=v_{a,L}=v_{a,0}$, $r_{d,a,H}<r_{d,a,0}<r_{d,a,L}$, which is termed fixed $v_a$ constellation, and 2) place the symbols on vertical line through $\mathbf{x}_{a,0}$ such that $r_{d,a,H}=r_{d,a,L}=r_{d,a,0}$, $v_{a,H}>v_{a,0}>v_{a,L}$, termed fixed $r_{d,a}$ constellation.
We evaluate the error probability performance for these two constellations in Section~\ref{sec:evaluation}.

\subsection{Receiver Design}

We focus on a simple and robust receiver that can cope with the uncertainties of the load variation by averaging the received DC level in each symbol interval:
\begin{equation}\label{eq.7}
\overline{y}_b=\frac{1}{N}\sum_{n=1}^N y_b[n]=\frac{1}{N}\sum_{n=1}^N(v^*[n]+z_b[n])=\overline{v^*}+\overline{z_b}\cong\overline{v^*},
\end{equation} 
where $y_b[n],n=1,...,N$ are samples of the received signal (\ref{eq.3}) and $N$ is the number of samples per symbol interval. \footnote{In practice the sampling frequency $f_s$ is typically of the order of $10$~kHz and the symbol duration $T_S$ of the order of 1--100~ms, which further justifies the assumption $\overline{z}\approx0$.}
The receiver uses simple hypothesis testing against the threshold obtained through the pilot:
\begin{equation}\label{eq.8}
\overline{y}\; \left\{
  \begin{array}{lr}
    > v_0^* & \text{decide} \; \mathbf{x}_{a,H},  \\
    < v_0^* & \text{decide} \; \mathbf{x}_{a,L}. 
  \end{array}
\right.
\end{equation}
Note that the receiver does not have to know the exact values of $v_{H}^*$ and $v_{L}^*$, and of the corresponding symbols $\mathbf{x}_{a,H}$ and $\mathbf{x}_{a,L}$,  to perform detection by hypothesis testing in \eqref{eq.8}.
Finally, as the transmitter observes the same bus voltage, the same detection mechanism can be employed at the transmitter side in order to deduce the symbol decisions made by the receiver and compare them to the symbols that are actually sent.


\subsection{Line Coding and Coping with Errors}

The receiver makes incorrect symbol decision whenever the load $r$ changes in a manner that the average DC level, computed by \eqref{eq.7}, is flipped at the other side of the threshold.
In other words, the erroneous symbol decisions are made when both $v_H^*$ and $v_L^*$ are on the same side of the threshold. 
Sections~\ref{sec:analysis} and \ref{sec:evaluation} analyze in detail the probability of symbol error, given that the load has changed during the symbol interval, illustrating the robustness of the proposed receiver (\ref{eq.7}) to load variations for different symbol constellations.
However, the major challenge is whether reliable communication is possible \emph{after} an error-inducing load change has occurred?
To this end, we introduce the probability of joint detection of an error-inducing load change by both the transmitter and the receiver, $p_{e}=\text{Pr}(e\;in\;TX,\;e\;in\;RX)$
and the probability of burst of errors after an an error-inducing load change has occurred, $p_{burst|e}$.
In sequel, we propose two simple protocols and discuss their error managing capabilities.

\begin{figure*}[b]
\normalsize
\hrulefill
\vspace*{4pt}
\setcounter{MYtempeqncnt}{\value{equation}}
\setcounter{equation}{24}
\begin{align}\label{eq.21}
p_{1\rightarrow0} & = \mathbb{E}_{R_0}\left\{\int\limits_{0}^TC_R\bigg(\bigg(\bigg(\frac{Tv_0^*}{\alpha_{H}(T-\theta)}-\frac{\theta}{(\beta_{H}+r_0^{-1})(T-\theta)}\bigg)^{-1}-\beta_{H}\bigg)^{-1}\bigg)p_{\Theta}(\theta|\theta<T)d\theta\right\}, \\
\label{eq.22}
p_{0\rightarrow1} & = \mathbb{E}_{R_0}\left\{\int\limits_{0}^T\bigg[1-C_R\bigg(\bigg(\bigg(\frac{Tv_0^*}{\alpha_{L}(T-\theta)}-\frac{\theta}{(\beta_{L}+r_0^{-1})(T-\theta)}\bigg)^{-1}-\beta_{L}\bigg)^{-1}\bigg)\bigg]p_{\Theta}(\theta|\theta<T)d\theta\right\} .
\end{align}
\setcounter{equation}{\value{MYtempeqncnt}}
\end{figure*}

The first scheme is a simple binary signaling: `1' is sent via $\mathbf{x}_{a,H}$ and `0' is sent via $\mathbf{x}_{a,L}$ in a single bit interval; thus, $T = T_S$.
In this case, the coding and detection strategies combined transform the MG bus into Binary Asymmetric Channel (BAC) with the bit flip (i.e., error) probabilities determined by the load behavior.
The induced communication channel is asymmetric because of the non-linear model \eqref{therip}; the corresponding analysis is performed in Section~\ref{sec:analysis}.
In absence of noise, if the load changes such that a bit flip occurs, the corresponding erroneous decision will be detected by the transmitter, but not by the receiver, implying that $p_{e}=0$.
Moreover, if the receiver continues to use the same threshold as prior to the load change, then, depending on the load change, either all `1's or all `0' will be flipped, resulting in a continuous stream of `1's or `0's and $p_{burst|e}=1$.
Such a scenario is depicted in Fig.~\ref{Fig6}.
To prevent this from happening necessitates the implementation of additional techniques for error detection and correction; the fact that the transmitter knows what the receiver has decided opens prospect to design elegant solutions in this respect.
A simple choice is to limit the length of consecutive all '1's and '0's sequences, so that if such a sequence occurs, the receiver concludes that the error event has taken place and awaits for the subsequent threshold reset.

\begin{figure}[t]
\centering
\includegraphics[scale=0.5]{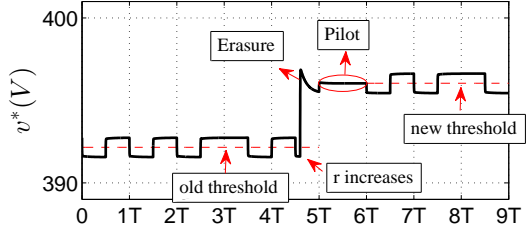}
\caption{Coping with errors in Manchester coding.}
\label{Fig9}
\end{figure}

The other strategy is Manchester coding on bit level using the following waveform: `1' is sent as $[\mathbf{x}_{a,H},\;\mathbf{x}_{a,L}]$ and `0' is transmitted as $[\mathbf{x}_{a,L},\;\mathbf{x}_{a,H}]$.
Thus, a single bit interval is composed of two symbol intervals, i.e., $T = 2 T_S$.
In this scheme both the transmitter and the receiver can perfectly and identically detect that an error-inducing load change has occurred if the average voltage levels in the first and the second symbol interval are on the same side of the threshold, i.e., $p_{e}=1$ (recall that we assume a maximum of one load change in a \emph{bit} interval).
In other words, the MG bus can be seen as a Binary Asymmetric Erasure Channel (BAEC) on bit level.
As both the transmitter and the receiver have common knowledge about the error-inducing event, they can perform coordinated actions to deal with it and prevent potential error burst. 
A simple and adequate strategy is to insert the pilot by the transmitter after error detection, resetting the detection threshold at the receiver, leading to $p_{burst|e}=0$.
The example of such protocol operation is depicted in Fig.~\ref{Fig9}. 
On the other hand, Manchester coding requires double bandwidth w.r.t. simple binary signaling, but it also enables achieving and maintaining symbol-level synchronization.

We conclude this section by noting that introducing noise in the system alters the performance of the proposed strategies and, in general, implies $0<p_{e}<1$.
In general, higher layer error correction codes can be used to deal with noise-related errors, which is out of the paper scope and left for future work.


\section{Analysis}
\label{sec:analysis}

This section develops framework for analytical modeling of the power-talk schemes proposed in Section~\ref{sec:binary}. 
The load $r$ can be modeled as a random variable, as shown in Fig.~\ref{Fig4}.
Let $R\sim p_R(r)$, where $p_R(r)$ is a general pdf with a finite support, i.e., $R\in[R_{min},\;R_{max}]$.
The cdf of $R$ is denoted with $C_R(r)=\text{Pr}(R\leq r)$.
We consider a MG with step load changes and assume that no more than one load change occurs during a bit interval and that a steady state is reached fast after a change; these assumptions are verified in practical MG setup.
The time instant of load change $\theta$ is distributed according to $p_{\Theta}(\theta)$, $\theta\geq0$, where 0 denotes the start of the bit interval.

The simple binary signaling transforms the MG bus in BAC in each bit interval.
Given the initial load $r_0$ and the change instant $\theta$, the conditional bit flip probabilities are:
\begin{align}\
 p_{1\rightarrow0} & (\mathbf{x}_{a,0},\mathbf{x}_{a,H}|r_0,\theta) = \text{Pr}(\overline{v}_{H,(0,T)}<v_0^*)\\
 \label{eq:p1_BAC}
 \cong \, & \, \text{Pr}\bigg(f(\mathbf{x}_{a,H},r_0)\frac{\theta}{T}+f(\mathbf{x}_{a,H},r)\frac{T-\theta}{T}<v_0^*\bigg)\\\label{eq.19}
 = \, & \text{Pr}\bigg(r<\frac{1}{(\frac{Tv_0^*}{\alpha_{H}(T-\theta)}-\frac{\theta}{(\beta_{H}+r_0^{-1})(T-\theta)})^{-1}-\beta_{H}}\bigg) , \\
p_{0\rightarrow1} & (\mathbf{x}_{a,0},\mathbf{x}_{a,L}|r_0,\theta)=\text{Pr}(\overline{v}_{L,(0,T)}>v_0^*)\\
\label{eq:p0_BAC}
 \cong \, & \text{Pr}\bigg(f(\mathbf{x}_{a,L},r_0)\frac{\theta}{T}+f(\mathbf{x}_{a,L},r)\frac{T-\theta}{T}<v_0^*\bigg)\\\label{eq.20}
= \, &  \text{Pr}\bigg(r>\frac{1}{(\frac{Tv_0^*}{\alpha_{L}(T-\theta)}-\frac{\theta}{(\beta_{L}+r_0^{-1})(T-\theta)})^{-1}-\beta_{L}}\bigg).
\end{align}
Eq.  \eqref{eq:p1_BAC}/\eqref{eq:p0_BAC} are obtained using the assumption of step load changes with negligible transient time, and  \eqref{eq.19}/\eqref{eq.20} using the steady-state MG model (\ref{therip}).
The bit flip probabilities, $p_{1\rightarrow0}$ and $p_{0\rightarrow1}$, are obtained through averaging for $\theta$ and $r_0$:
\begin{align}\label{eq.17}
p_{1\rightarrow0}=\mathbb{E}_{R_0}\left\{\mathbb{E}_{\Theta|\Theta<T}\left\{p_{1\rightarrow0}(\mathbf{x}_{a,0},\mathbf{x}_{a,H}|r_0,\theta)\right\}\right\},\\
\label{eq.18}
p_{0\rightarrow1}=\mathbb{E}_{R_0}\left\{\mathbb{E}_{\Theta|\Theta<T}\left\{p_{0\rightarrow1}(\mathbf{x}_{a,0},\mathbf{x}_{a,L}|r_0,\theta)\right\}\right\},
\end{align}
where $\mathbb{E}_{\Theta|\Theta<T}\left\{.\right\}$ is performed w.r.t. the conditional pdf $p_{\Theta}(\theta|\theta<T)$.
The bit flip probabilities $p_{1\rightarrow0}$ and $p_{0\rightarrow1}$ are given in the general form in (\ref{eq.21}) and (\ref{eq.22}).
\addtocounter{equation}{2}

\begin{figure*}[t]
\normalsize
\setcounter{MYtempeqncnt}{\value{equation}}
\setcounter{equation}{30}
\begin{align}\nonumber
q_{1\rightarrow e} = & \mathbb{E}_{R_0}\left\{\int\limits_{0}^{T/2}\bigg[1-C_R\bigg(\bigg(\frac{\alpha_{L}}{v_0^*}-\beta_{L}\bigg)^{-1}\bigg)+C_R\bigg(\bigg(\bigg(\frac{\frac{T}{2}v_0^*}{\alpha_{H}(\frac{T}{2}-\theta)}-\frac{\theta}{(\beta_{H}+r_0^{-1})(\frac{T}{2}-\theta)}\bigg)^{-1}-\beta_{H}\bigg)^{-1}\bigg)\bigg]p_{\Theta}(\theta|\theta<T)d\theta\right\}\\\label{eq.28}
& + \mathbb{E}_{R_0}\left\{\int\limits_{T/2}^{T}\bigg[1-C_R\bigg(\bigg(\bigg(\frac{\frac{T}{2}v_0^*}{\alpha_{L}(T-\theta)}-\frac{\theta-\frac{T}{2}}{(\beta_{L}+r_0^{-1})(T-\theta)}\bigg)^{-1}-\beta_{L}\bigg)^{-1}\bigg)\bigg]p_{\Theta}(\theta|\theta<T)d\theta\right\}, \\
\nonumber
q_{0\rightarrow e} = & \mathbb{E}_{R_0}\left\{\int\limits_{0}^{T/2}\bigg[1-C_R\bigg(\bigg(\bigg(\frac{\frac{T}{2}v_0^*}{\alpha_{L}(\frac{T}{2}-\theta)}-\frac{\theta}{(\beta_{L}+r_0^{-1})(\frac{T}{2}-\theta)}\bigg)^{-1}-\beta_{L}\bigg)^{-1}\bigg)+C_R\bigg(\bigg(\frac{\alpha_{H}}{v_0^*}-\beta_{H}\bigg)^{-1}\bigg)\bigg]p_{\Theta}(\theta|\theta<T)d\theta\right\}\\\label{eq.29}
& + \mathbb{E}_{R_0}\left\{\int\limits_{T/2}^{T}C_R\bigg(\bigg(\bigg(\frac{\frac{T}{2}v_0^*}{\alpha_{H}(T-\theta)}-\frac{\theta-\frac{T}{2}}{(\beta_{H}+r_0^{-1})(T-\theta)}\bigg)^{-1}-\beta_{H}\bigg)^{-1}\bigg)p_{\Theta}(\theta|\theta<T)d\theta\right\} .
\end{align}
\setcounter{equation}{\value{MYtempeqncnt}}
\hrulefill
\end{figure*}

\begin{figure*}[b]
\centering
\subfloat[Simple binary signaling: $1\rightarrow 0$.]{\includegraphics[scale=0.3]{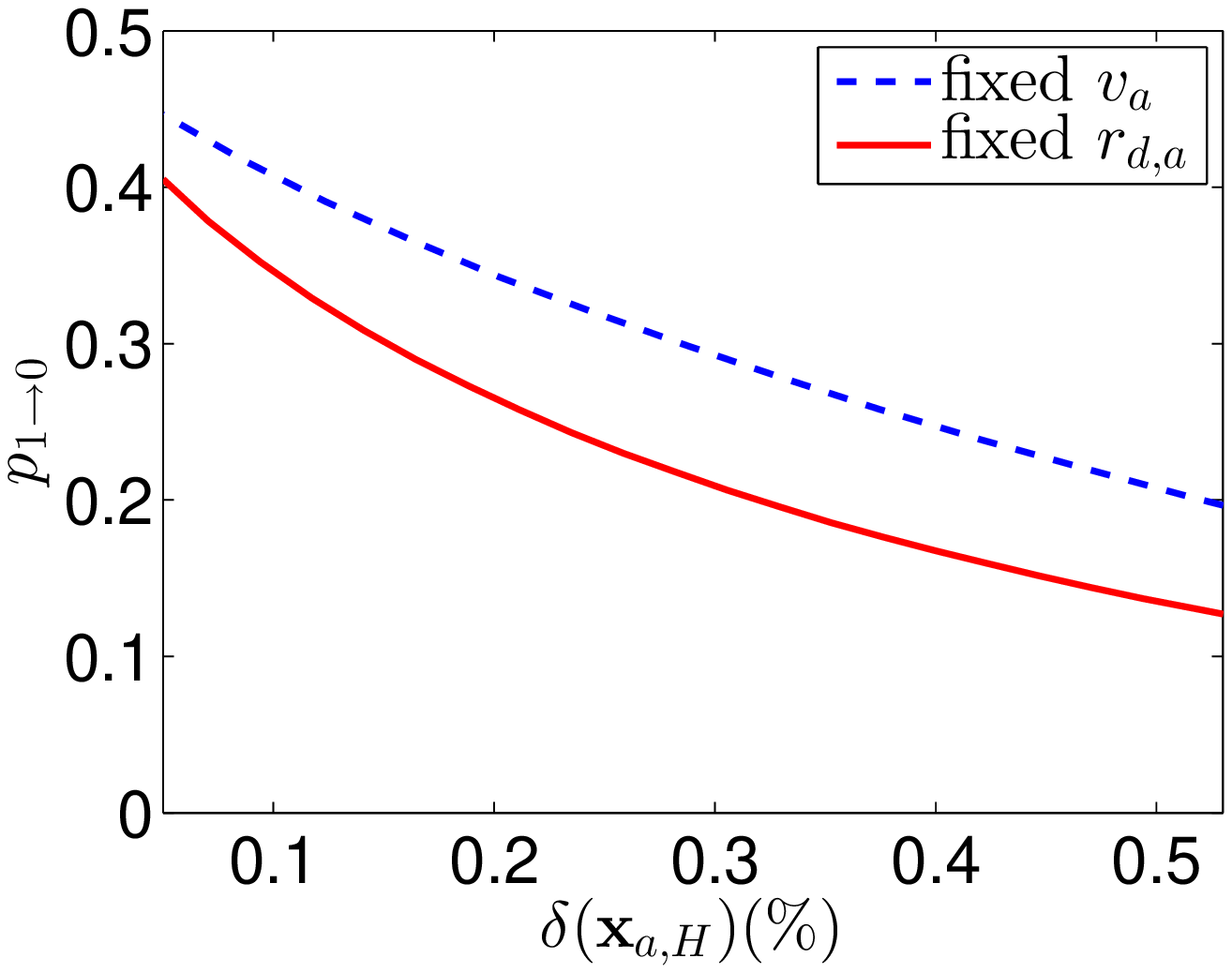} \label{10a}}
\hfil
\subfloat[Simple binary signaling: $0\rightarrow 1$.]{\includegraphics[scale=0.3]{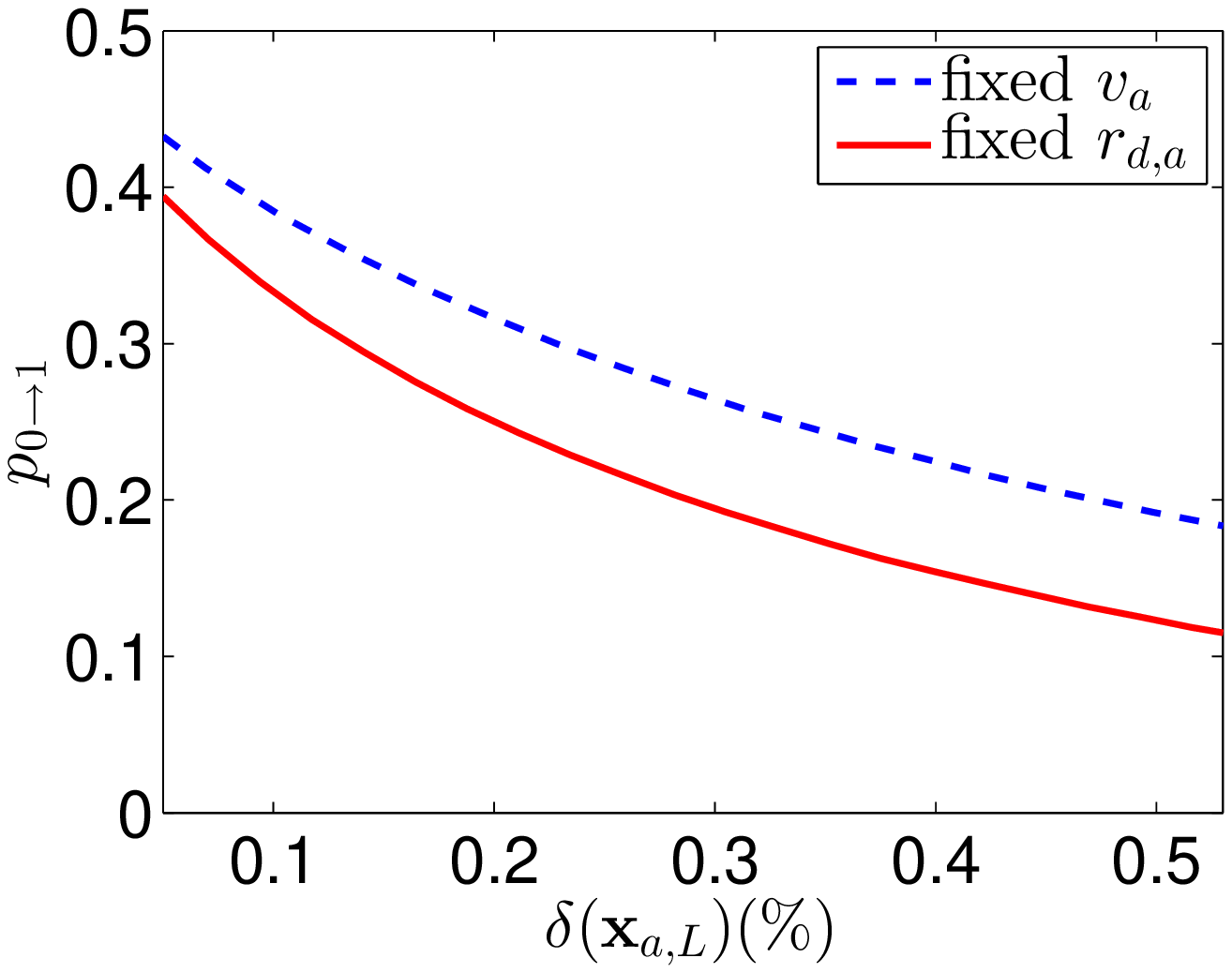} \label{10b}}
\hfil
\subfloat[Manchester coding: $1\rightarrow e$.]{\includegraphics[scale=0.3]{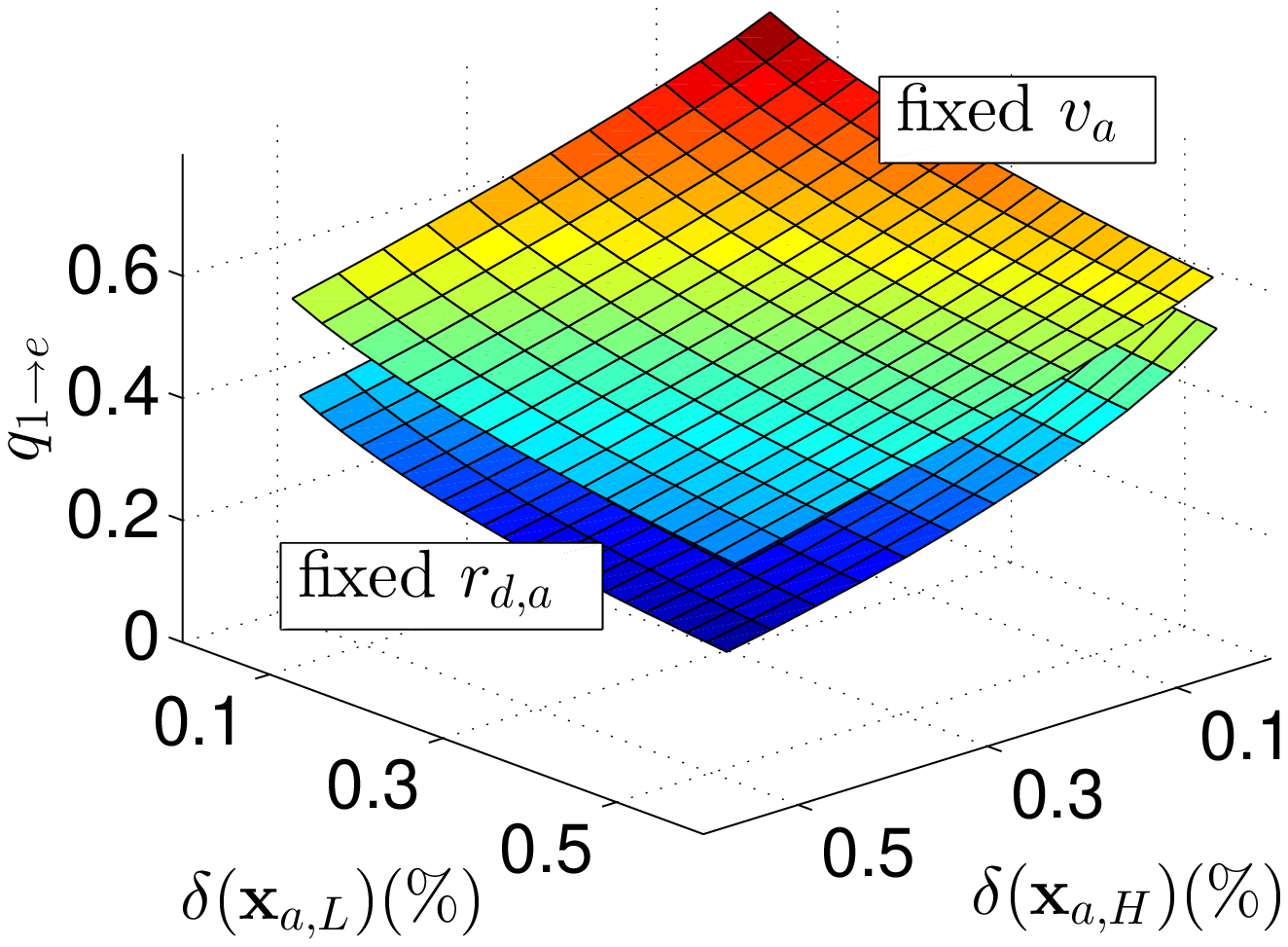} \label{10c}}
\hfil
\subfloat[Manchester coding: $0\rightarrow e$.]{\includegraphics[scale=0.3]{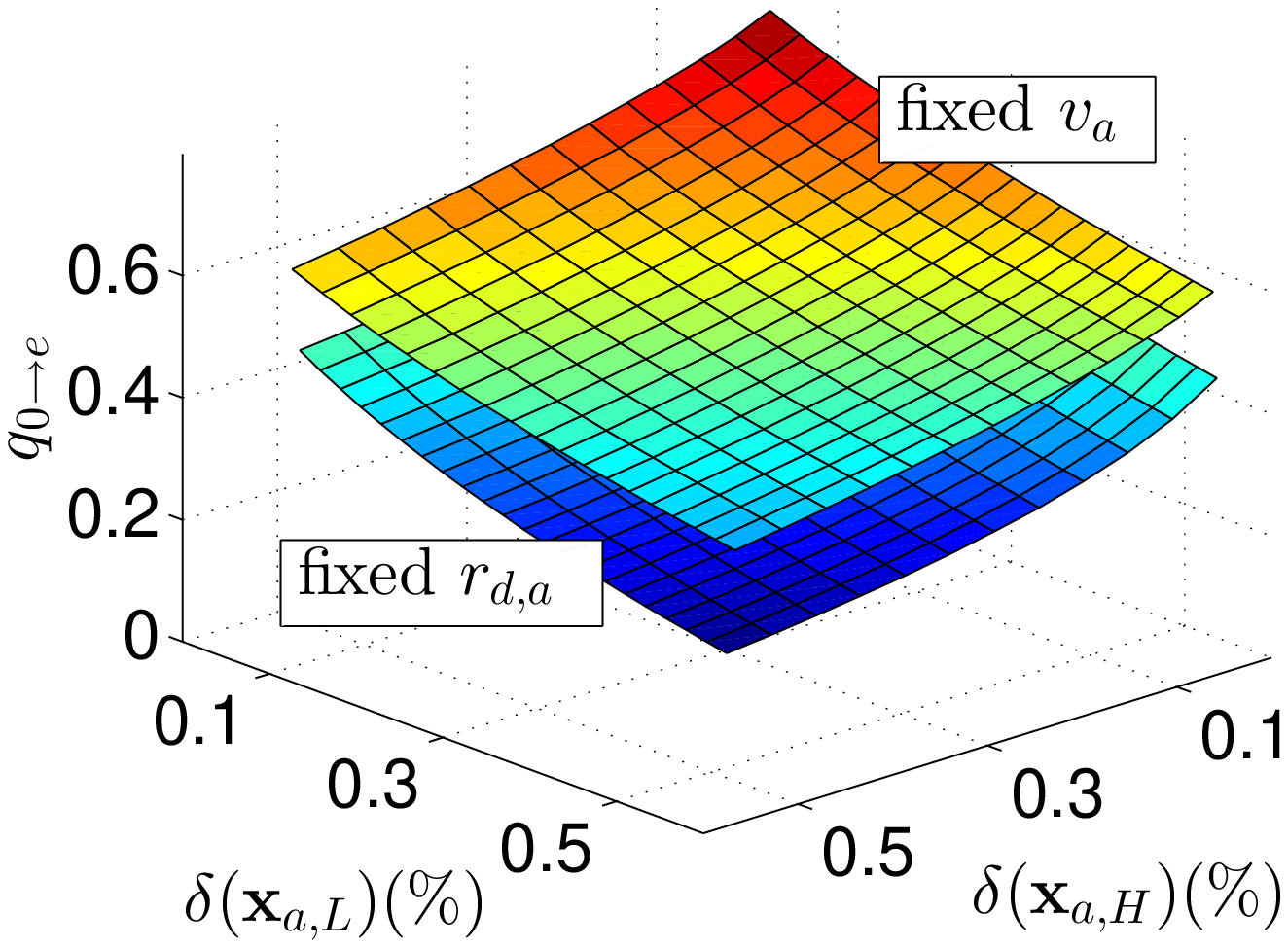} \label{10d}}
\caption{The bit error probability given that the load has changed.}
\label{Fig10}
\end{figure*}

We now analyze the case when Manchester coding is used.
In absence of noise and given that the load has changed only once during the bit interval, the MG bus is transformed into BAEC.
The conditional erasure probabilities are:
\begin{align}\nonumber
 q_{1\rightarrow e}  (\mathbf{x}_{a,0},  \mathbf{x}_{a,H}, & \mathbf{x}_{a,L}|  r_0,\theta)  = \\
\label{eq.26}
& \text{Pr}(\overline{v}_{H,(0,\frac{T}{2})}>v_0^*,\overline{v}_{L,(\frac{T}{2},T)}>v_0^*) \,+ \nonumber \\
& \text{Pr}(\overline{v}_{H,(0,\frac{T}{2})}<v_0^*,\overline{v}_{L,(\frac{T}{2},T)}<v_0^*), \\
\nonumber
 q_{0\rightarrow e}  (\mathbf{x}_{a,0}, \mathbf{x}_{a,H}, & \mathbf{x}_{a,L}| r_0,\theta) \nonumber = \\
\label{eq.27}
& \text{Pr}(\overline{v}_{L,(0,\frac{T}{2})}>v_0^*,\overline{v}_{H,(\frac{T}{2},T)}>v_0^*) \, + \nonumber \\
& \text{Pr}(\overline{v}_{L,(0,\frac{T}{2})}<v_0^*,\overline{v}_{H,(\frac{T}{2},T)}<v_0^*).
\end{align}
We illustrate the derivation for $q_{1\rightarrow e}(\mathbf{x}_{a,0},\mathbf{x}_{a,H},\mathbf{x}_{a,L}|r_0,\theta)$, noting that the same principles apply to the derivation of $q_{0\rightarrow e}(\mathbf{x}_{a,0},\mathbf{x}_{a,H},\mathbf{x}_{a,L}|r_0,\theta)$.
The load change can occur either in the first or the second half of the bit interval (i.e., in the first or second symbol interval); using the assumption of step change with negligible transient time, we have:
\begin{align}\nonumber
& \text{Pr}(\overline{v}_{H,(0,\frac{T}{2})}>v_0^*,\overline{v}_{L,(\frac{T}{2},T)}>v_0^*) \cong \\\nonumber
& \text{Pr}\bigg(f(\mathbf{x}_{a,H},r_0)\frac{\theta}{\frac{T}{2}}+f(\mathbf{x}_{a,H},r)\frac{\frac{T}{2}-\theta}{\frac{T}{2}}>v_0^*, f(\mathbf{x}_{a,L},r)>v_0^*\bigg)\\
\nonumber
& +\text{Pr}\bigg(f(\mathbf{x}_{a,H},r_0)>v_0^*, f(\mathbf{x}_{a,L},r_0)\frac{\theta-\frac{T}{2}}{\frac{T}{2}} \, +  \\
&  f(\mathbf{x}_{a,L},r)\frac{T-\theta}{\frac{T}{2}}>v_0^*\bigg), \label{new1}\\
\nonumber
& \text{Pr}(\overline{v}_{H,(0,\frac{T}{2})}<v_0^*,\overline{v}_{L,(\frac{T}{2},T)}<v_0^*) \cong \\\nonumber
& \text{Pr}\bigg(f(\mathbf{x}_{a,H},r_0)\frac{\theta}{\frac{T}{2}}+f(\mathbf{x}_{a,H},r)\frac{\frac{T}{2}-\theta}{\frac{T}{2}}<v_0^*, f(\mathbf{x}_{a,L},r)<v_0^*\bigg) \\ \nonumber
& +\text{Pr}\bigg( f(\mathbf{x}_{a,H},r_0)<v_0^*,  f(\mathbf{x}_{a,L},r_0)\frac{\theta-\frac{T}{2}}{\frac{T}{2}} \, +\\
\label{new2} & f(\mathbf{x}_{a,L},r)\frac{T-\theta}{\frac{T}{2}}<v_0^*\bigg).
\end{align}
Using the standard formula $\text{Pr}(A,B)=\text{Pr}(A|B)\text{Pr}(B)$, it can be shown that both joint probabilities in (\ref{new1}) are equal to the probabilities of their respective second events.
Further, the first joint probability in (\ref{new2}) is equal to the probability of the first event and the second joint probability in (\ref{new2}) is 0.
Using the model (\ref{therip}) and after rearranging (\ref{new1}) and (\ref{new2}) leads to (\ref{eq.28}) for $q_{1\rightarrow 0}$; applying the same approach for $q_{0\rightarrow e}(\mathbf{x}_{a,0},\mathbf{x}_{a,H},\mathbf{x}_{a,L}|r_0,\theta)$, gives (\ref{eq.29}) for $q_{0\rightarrow e}$.

We conclude the analysis by remarks related to the inclusion of the effects of the noise and multiple load changes per bit interval.
For simple binary signaling, the channel will remain BAC, but the calculation of the bit flip probabilities will change.
For Manchester coding, however, the bit flip probabilities will be non-zero when accounting for the noise and multiple load changes.
In this case, the resulting channel will be a combination of BAEC and BAC.
Our preliminary investigations show that the erasures are still the dominant error events by far, since the noise is small and the occurrence of multiple load changes within bit interval unlikely.

\section{Performance Evaluation}
\label{sec:evaluation}

We evaluate the performance of the proposed communication schemes using the following MG parameters: $V_{min}=390 \, \text{V}$, $V_{max}=400 \, \text{V}$, $I_{a,max}=6 \, \text{A}$, $I_{b,max}=4 \, \text{A}$, $v_b=400 \, \text{V}$, $r_{d,b}=\frac{v_b-V_{min}}{I_{b,max}}$, $R_{min}=50 \, \Omega$, $R_{max}=250 \, \Omega$.
The signaling space for these values is shown in Fig.~\ref{Fig5}.
The pilot symbol is $\mathbf{x}_{a,0}=(v_{a,0},\frac{v_{a,0}-V_{min}}{I_{a,max}})$ where $v_{a,0}=400 \, \text{V}$.
We describe the load distribution using a non-informative prior, i.e., assuming the uniform distribution $R\sim\mathcal{U}[R_{min},R_{max}]$; this scenario can be considered as the worst case.
Also, we assume an exponential distribution for $\theta$, i.e.,  $\theta\sim Exp(T^{-1})$.


We illustrate the error probabilities for the two symbol constellations shown in Fig.~\ref{Fig7}, i.e., (i) the fixed $v_a$ constellation, where $1.67 \, \Omega>r_{d,a,H}>0.8 \, \Omega$ and $1.67 \, \Omega<r_{d,a,L}<2.5 \, \Omega$ and (ii) fixed $r_{d,a}$ constellation, where $400 \, \text{V}<v_{a,H}<402 \, \text{V}$ and $400\, \text{V}>v_{a,L}>396 \, \text{V}$.
Fig.~\ref{Fig10}\subref{10a} and Fig.~\ref{Fig10}\subref{10b} show the bit flip probabilities for the simple binary signaling, and Fig.~\ref{Fig10}\subref{10c} and Fig.~\ref{Fig10}\subref{10d} show the erasure probabilities for Manchester coding, as functions of the average power deviation (\ref{latest1}) in \%, for each symbol (either $\mathbf{x}_{a,H}$ or $\mathbf{x}_{a,L}$) relative to the pilot.
Evidently, the error probabilities for both channels decrease as $\delta$ increases due to better separation between the signaling levels $v_L^*$, $v_H^*$ and the threshold level $v_0^*$, demonstrating a fundamental trade-off between the performance of the power talk and the disturbances it induces in the system, as expressed through the power deviation metric.
The results also reveal that for the same power deviation, using fixed $r_{d,a}$ constellation is a better choice than using fixed $v_a$ constellation, since it demonstrates better error probability performance both for simple binary and Manchester coding.
In other words, separating the symbols along the referent voltage dimension in the signaling space, see Fig.~\ref{Fig7}, leads to more reliable channels.
This observation is also in direct relation with the non-isotropic behavior of $\delta(\mathbf{x}_a)$ as function of the distance $d$ (illustrated on Fig.~\ref{Fig_diss}), suggesting that for given system settings, the power deviation can be minimized by placing the symbols in particular directions in the signaling space. The design of such optimal symbol constellations is left for future work.

\begin{figure}[t]
\centering
\includegraphics[scale=0.5]{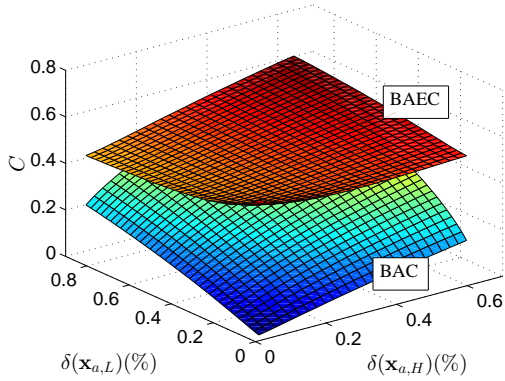}
\caption{The conditional capacity per bit interval $C$~[bit/bit~interval], provided that the load changed in the interval (fixed $r_{d,a}$ constellation).}
\label{Fig12}
\vspace{-12pt}
\end{figure}

Finally, we compare the performance of the proposed power-talk schemes in terms of the conditional capacity per bit interval given that the load has changed, $C$ [bit/bit interval].\footnote{Note that the average capacity per bit interval can be calculated as $\overline{C}\text{ [bit/bit interval]} = C \, p_c +1 (1- p_c )$, where $p_c$ is the probability of a load change in the bit interval and where 1~[bit/bit~interval] is the capacity of the binary power talk given that load has not changed. If divided by the length of the bit interval $T$, the average capacity can be expressed in bps.}
Since the respective channels are asymmetric, we use Arimoto-Blahut algorithm \cite{ref11} to evaluate their respective conditional capacities.
Fig. \ref{Fig12} shows the results only for the fixed $r_{d,a}$ constellation due to space limitation; we note that the same trends are observed for the fixed $v_a$ constellation.
Evidently, the capacity improvement of the BAEC over the BAC is substantial, especially in the region of small $\delta$, which, as already discussed, is of great practical importance, which is 
yet another reason to use the Manchester coding as a more reliable choice.



\section{Discussion and Conclusions}
\label{sec:conclusions}

In this paper, we introduced the power talk - a novel communication strategy for MGs where the communication is carried out by modulating the power signals. 
Specifically, the proposed instance of the power talk exploits the inherently available bus parameter deviation space and modulates information in the variations of the bus voltage.
The strategy is challenged by the random variations of the voltage level, as a result of load changes in the system.
We show that it is possible to communicate reliably over the MG using common schemes, such as simple binary signaling or Manchester coding.

The achievable signaling rates of the power talk are determined by the symbol duration $T_S$.
In practice, $T_S$ should be of the order of 1--100~ms to allow the MG bus to reach a steady state within a symbol interval, leading to the rate of the order of 0.01--1 kBaud.
Although such a rate seems low, it is likely to be sufficient for all purposes of local MG management.
Moreover, power talk offers a high reliability and availability, equal to the reliability and availability of the MG bus itself.

The proposed concept opens multiple directions for future research, such as: design of higher-order constellations, two- and multi-way communication, inclusion of the effects of the transmission network and different types of loads, such as constant power and slowly changing loads, etc.


\section*{Acknowledgement}

The work presented in this paper was supported in part by EU, under grant agreements no. 607774 ``ADVANTAGE'' and no. 619437 ``SUNSEED'', and in part by the Danish Council for Independent Research (DFF), grant no. 4005-0028.

\bibliographystyle{IEEEtran}

\end{document}